\documentclass[12pt]{article}
\usepackage{amsmath}
\usepackage{bm}
\usepackage{color}
\usepackage[T1]{fontenc}
\usepackage{graphicx}
\usepackage{caption}
\usepackage{subcaption}
\begin{document}
\begin{center}
\begin{large}
{\bf Entanglement of graph states of spin system with Ising interaction and its quantifying on IBM's quantum computer}
\end{large}
\end{center}

\centerline {Kh. P. Gnatenko \footnote{E-Mail address: khrystyna.gnatenko@gmail.com} V. M. Tkachuk \footnote{E-Mail address: voltkachuk@gmail.com}}
\medskip
\centerline {\small \it Ivan Franko National University of Lviv, Department for Theoretical Physics,}
\centerline {\small \it 12 Drahomanov St., Lviv, 79005, Ukraine}

\begin{abstract} We consider graph states generated by operator of evolution with Ising Hamiltonian. The geometric measure of entanglement of a spin with other spins in the graph state is obtained analytically and quantified on IBM's quantum computer, IBM Q Valencia.  The results of quantum calculations are in good agreement with the theoretical ones.  We conclude that the geometric measure of entanglement of a spin with other spins in the graph state is related with degree of vertex representing the spin in the corresponding graph.
\end{abstract}

\section{Introduction}
Entanglement is a critical resource in quantum communications and quantum computing (see, for instance, \cite{Feynman,Horodecki,Raussendorf,Lloyd,Shi,Llewellyn,Huang,Yin,Jennewein,Karlsson} and references therein). Its calculation plays important role in quantum information. Therefore much attention has been devoted to studies of entanglement of quantum states and its detecting on quantum computers \cite{Horodecki,Shimony,Behera,Scott,Horodecki1,Samar,Kuzmak,Kuzmak1,Wang,Mooney}.

The geometric measure of entanglement is defined as minimal squared Fubiny-Study distance between an entangled state $\vert\psi\rangle$ and a set of separable pure states $\vert\psi_s\rangle$, namely it reads $E(\vert\psi\rangle)=\min_{\vert\psi_s\rangle}(1-|\langle\psi|\psi_s\rangle|^2)$. This measure of entanglement was proposed by Shimony \cite{Shimony}.
In the paper \cite{Samar} it was found that the geometric measure of entanglement of a spin with a quantum system in pure state
\begin{eqnarray}
\mid\psi\rangle=a\vert\uparrow\rangle\vert\Phi_1\rangle+b\vert\downarrow\rangle\vert\Phi_2\rangle,\label{pure}
 \end{eqnarray}
(here $a$, $b$ are real and positive constants, $\vert\Phi_1\rangle$, $\vert\Phi_2\rangle$ are arbitrary state vectors of a quantum system with norm equals to one, $\langle\Phi_i\vert\Phi_i\rangle=1$, $i=1,2$)
is related with the mean value of the spin.
Namely, the entanglement of a spin-1/2 with a quantum system  in pure state (\ref{pure})
reads
\begin{eqnarray}
E(\vert\psi\rangle)=\frac{1}{2}(1-|\langle{\bm \sigma}\rangle|),\label{ent}
\end{eqnarray}
where $|\langle{\bm \sigma}\rangle|=\sqrt{\langle{\bm \sigma}\rangle}$, the components $\sigma^x$, $\sigma^y$, $\sigma^z$ of vector $\bm{ \sigma}$  are the Pauli matrixes. Therefore, measuring the mean value of spin one can detect the geometric measure of entanglement.

Much attention has been devoted to studies of graph states (see, for instance, \cite{Wang,Mooney,Hein,Hein1,Guhne,Markham,Cabello,Alba,Mezher} and references therein).
Well studied are graph states generated by 2-qubit controlled-Z operator (see, for instance, \cite{Wang,Mooney,Cabello,Alba,Mezher}). In recent paper \cite{Wang} graph states that correspond to  rings were prepared on 16-qubit IBM's quantum computer. It was shown that 16-qubit IBM's quantum computer (ibmqx5) can be fully entangled. The authors of paper \cite{Mooney} prepared graph state on the 20-qubit quantum device IBM Q Poughkeepsie and examined its entanglement.

 In the present paper we consider graph states of spin systems generated by operator of evolution  with Ising Hamiltonian. Such generation of the graph states opens possibility to consider them as states of physical systems. We find expression for geometric measure of entanglement of a spin with other spins in the graph states and conclude that the entanglement depends on  the number of edges that are incident to the vertex representing the spin. The geometric measure of entanglement of a spin with other spins in the graph state was also quantified on the IBM's quantum computer.

The paper is organized as follows. In the Section 2 we find analytically the geometric measure of entanglement of a spin with other spins in the graph states  generated by operator of evolution  with Ising Hamiltonian. The relation of the the geometric measure of entanglement with graph  properties is examined. Section 3 is devoted to studies of the geometric measure of entanglement of the graph states on IBM's quantum computer. We present the results of quantifying entanglement of a spin with other spins in the graph state on 5-qubit IBM's quantum computer (IBM Q Valencia). Conclusions are presented in Section 4.

\section{Geometric measure of entanglement of graph states}

Let us study graph states generated by operator of evolution with Ising Hamiltonian.
We consider a system of $N$ spins which is described by the Hamiltonian
 \begin{eqnarray}
H=\frac{1}{2}\sum_{ij}J_{ij}\sigma_i^x\sigma_j^x,\label{hamilt}
\end{eqnarray}
here $\sigma_i^x$ is the Pauli matrix of spin $i$,  $J_{ij}$  is the interaction coupling, $i,j=1..N$.

Starting from the initial state
\begin{eqnarray}
\vert\psi_0\rangle=\vert00...0\rangle,\label{zero}
\end{eqnarray}
(the spin states can be associated with qubit states, the state $|\uparrow\rangle$ corresponds to $|0\rangle$ and  $|\downarrow\rangle$ corresponds to $|1\rangle$) in result of evolution one obtains the state
 \begin{eqnarray}
\vert\psi\rangle=e^{-\frac{it}{2\hbar}\sum_{ij}J_{ij}\sigma_i^x\sigma_j^x}\vert\psi_0\rangle,\label{state}
\end{eqnarray}
which can be associated with a graph $G(V,E)$.  The vertices in the graph $V$ represent the spins in the system. The  edges between the vertexes $E$ describe interaction between the spins.
 We consider $J_{ij} = J,$
 and associate $J_{ij}$ with elements of adjacency matrix of the undirected graph $A_{ij}$ ($A_{ij}=1$ if the interaction between spin $i$ and spin $j$ exists, and $A_{ij}=0$ if spin spin $i$ and spin $j$ do not interact, $A_{ij}=J_{ij}/J$).

Let us calculate the geometric measure of entanglement of one spin with the rest
spins of the system with Hamiltonian (\ref{hamilt}). According to (\ref{ent}) the geometric measure of entanglement is related with the mean value of the spin. So, let us consider the spin with index $l$ and find
$\langle{\bm \sigma}_l\rangle$ in the state (\ref{state}).
We have
\begin{eqnarray}
\langle{\sigma}^x_l\rangle=\langle\psi_0 \mid e^{\frac{it}{2\hbar}\sum_{ij}J_{ij}\sigma_i^x\sigma_j^x}{\sigma}^x_l  e^{-\frac{it}{2\hbar}\sum_{ij}J_{ij}\sigma_i^x\sigma_j^x}\mid\psi_0\rangle=\langle\psi_0 \mid{\sigma}^x_l\mid\psi_0\rangle=0.
\end{eqnarray}
The mean value of ${\sigma}^y_l$ in state (\ref{state}) reads
\begin{eqnarray}
\langle{\sigma}^y_l\rangle=\langle\psi_0 \vert e^{\frac{it}{2\hbar}\sum_{ij}J_{ij}\sigma_i^x\sigma_j^x}{\sigma}^y_l  e^{-\frac{it}{2\hbar}\sum_{ij}J_{ij}\sigma_i^x\sigma_j^x}\vert\psi_0\rangle=\nonumber\\=\langle\psi_0 \vert e^{\frac{it}{\hbar}\sum_{j}J_{jl}\sigma_j^x\sigma_l^x}{\sigma}^y_l  e^{-\frac{it}{\hbar}\sum_{j}J_{jl}\sigma_j^x\sigma_l^x}\vert\psi_0\rangle=
\langle\psi_0 \vert e^{\frac{i2t}{\hbar}\sum_{j}J_{jl}\sigma_l^x\sigma_j^x}{\sigma}^y_l\vert\psi_0\rangle=0,
\end{eqnarray}
where we use the identity
\begin{eqnarray}
e^{\frac{it}{\hbar}\sum_{j}J_{jl}\sigma_j^x\sigma_l^x}{\sigma}^y_l  e^{-\frac{it}{\hbar}\sum_{j}J_{jl}\sigma_j^x\sigma_l^x}=e^{\frac{i2t}{\hbar}\sum_{j}J_{jl}\sigma_l^x\sigma_j^x}{\sigma}^y_l,
\end{eqnarray}
which follows from the fact that ${\sigma}^y_l$ and $\sigma_l^x$ anticommute. Similarly for $\langle{\sigma}^z_l\rangle$ we have
\begin{eqnarray}
\langle{\sigma}^z_l\rangle=\langle\psi_0 \vert e^{\frac{it}{2\hbar}\sum_{ij}J_{ij}\sigma_i^x\sigma_j^x}{\sigma}^z_l  e^{-\frac{it}{2\hbar}\sum_{ij}J_{ij}\sigma_i^x\sigma_j^x}\vert\psi_0\rangle=
\langle\psi_0 \vert e^{\frac{i2t}{\hbar}\sum_{j}J_{lj}\sigma_l^x\sigma_j^x}{\sigma}^z_l\vert\psi_0\rangle=\nonumber\\=\langle\psi_0 \vert e^{\frac{i2t}{\hbar}\sum_{j}J_{lj}\sigma_l^x\sigma_j^x}\vert\psi_0\rangle.
\end{eqnarray}
Taking into account that $J_{ij} = J,$ and using the following notation
\begin{eqnarray}
\varphi=\frac{2Jt}{\hbar},\label{varphi}
\end{eqnarray}
one obtains
\begin{eqnarray}
\langle{\sigma}^z_l\rangle=\cos^{k_l}\varphi,
\end{eqnarray}
where $k_l$ is defined as
\begin{eqnarray}
 k_l=\sum_j \frac{J_{lj}}{J}\label{kl}
\end{eqnarray}

Using  (\ref{ent}), the geometric measure of entanglement of spin $l$ with other spins in the system has the following form
\begin{eqnarray}
E_l(\vert\psi\rangle)=\frac{1}{2}(1-|\cos^{k_l}\varphi|)\label{res}
\end{eqnarray}

Note that the entanglement (\ref{res})) is the same  in the case of antiferromagnetic interaction ($J>0$) and in the case of ferromagnetic interaction ($J<0$).
 It is also important to mention that $k_l$ given by (\ref{kl}) is the degree of the vertex $l$ in the graph. So, the geometric measure of entanglement of spin $l$   with other spins in the graph state (\ref{state}) is related with the number of edges that are incident to the vertex that represents the spin.

\section{Quantifying geometric measure of entanglement of graph states on IBM's quantum computer}

\begin{figure}[h!]
	\centering
	\includegraphics[width=8cm]{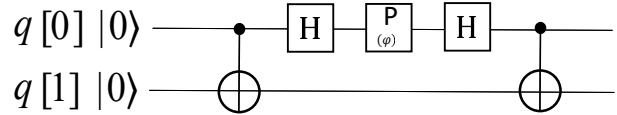}
	\caption{\footnotesize{Quantum protocol for preparing state (\ref{state1}). In the protocol controlled-NOT gate, Hadamard gate ($H$), Phase gate ($P(\varphi)$) are used, $\varphi$ is given by (\ref{varphi}).}}
	\label{p1}
\end{figure}

To prepare graph state (\ref{state}) on a quantum computer one has to realize action of operators
$\exp(-it J\sigma_i^x\sigma_j^x / {\hbar})$ on the state (\ref{zero}).  Note that operator $\exp(-it J\sigma_i^x\sigma_j^x / {\hbar})$ with exactness to total phase factor can be represented as $CX_{ij}H_iP_i(2Jt/\hbar)H_iCX_{ij}$ where $CX_{ij}$ is the controlled-NOT gate that acts on qubit labeled by index $i$ (q[i])
as ''control'' and on the qubit labeled by index $j$ ($q[j]$) as ''target''. Gate $H_i$ is the Hadamard gate acting on the $q[i]$, and $P_i(2Jt/\hbar)$ is the Phase gate which acts on $q[i]$ (phase gate does not change the state $\vert0\rangle$ and applies phase multiplier $\exp(i2Jt/\hbar)$ to the state $\vert1\rangle$). Or equivalently, operator $exp(-it J\sigma_i^x\sigma_j^x / {\hbar})$ can be represented as $CX_{ji}H_jP_j(2Jt/\hbar)H_jCX_{ji}$, where the controlled-NOT gate $CX_{ji}$  acts on qubit $q[j]$
as ''control'' and on the qubit $q[i]$ as ''target'', gates $H_j$, $P_j$ act on $q[j]$. For instance, quantum protocol for preparing two qubits $q[0]$ and $q[1]$ in the state
\begin{eqnarray}
\vert\psi\rangle=e^{-\frac{it}{\hbar}J\sigma_0^x\sigma_1^x}\vert00\rangle,\label{state1}
\end{eqnarray}
is presented on Fig. \ref{p1}
Using protocol Fig. \ref{p1} one can prepare states (\ref{state}) associated with different graphs.

  According to relation (\ref{ent}) for quantifying geometric measure of entanglement of spin $l$ with other spins in state (\ref{state}) on quantum computer the mean values of Pauli operators $\sigma_l^x$, $\sigma_l^y$, $\sigma_l^z$ in the state (\ref{state}) have to be measured.  The protocol for measuring these mean values was proposed in \cite{Kuzmak}. In the paper it was shown that the mean value of operators $\sigma^x$, $\sigma^y$, $\sigma^z$ can be represented by probabilities which define the result of measure on basis $\vert0\rangle$, $\vert1\rangle$. Namely, these mean values can be represented as
   \begin{eqnarray}
   \langle\sigma^x \rangle=\langle \psi \vert \sigma^x\vert\psi\rangle=\langle \tilde\psi^y \vert \sigma^z\vert \tilde\psi^y\rangle=\vert \langle \tilde\psi^y \vert 0 \rangle \vert^2-\vert \langle \tilde\psi^y \vert 1 \rangle \vert^2,\\
      \langle\sigma^y \rangle=\langle \psi \vert \sigma^y\vert\psi\rangle=\langle \tilde\psi^x \vert \sigma^z\vert \tilde\psi^x\rangle=\vert \langle \tilde\psi^x \vert 0 \rangle \vert^2-\vert \langle \tilde\psi^x \vert 1 \rangle \vert^2,\\
   \langle\sigma^z \rangle=\langle \psi \vert \sigma^z\vert\psi\rangle=\vert \langle\psi\vert 0 \rangle \vert^2-\vert \langle \psi \vert 1 \rangle \vert^2,
    \end{eqnarray}
where $\vert\tilde\psi^x\rangle=\exp(-i\pi\sigma^x/4)\vert\psi\rangle$,   $\vert\tilde\psi^y\rangle=\exp(-i\pi\sigma^y/4)\vert\psi\rangle$. From these relations follows that in order to measure the mean value $\langle \sigma^x_l \rangle$ before measurement in the standard basis one has to rotate the state of qubit $l$ around the $y$ axis by $\pi/2$ (one has to apply $RY(\pi/2)$ gate). In order to measure the mean value $\langle \sigma^y_l \rangle$ before measurement in the standard basis one has to rotate the state of qubit $l$ by $\pi/2$ around the $x$ axis (one has to apply $RX(\pi/2)$ gate).

We prepare graph state (\ref{state}) and quantify the geometric measure of entanglement on IBM's quantum computer  IBM Q Valencia.
IBM provides free access to this 5-qubit devise \cite{kk}. The 5 qubits interacts as it is shown in Fig. 1. Arrows indicate qubits between which the CNOT gate can
be directly applied (see Fig. 1).
\begin{figure}[h!]
	\centering
	\includegraphics[width=4cm]{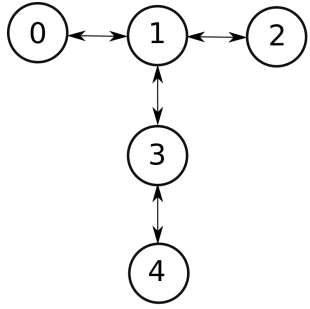}
	\caption{\footnotesize{Structure of IBM Q Valencia. Arrows indicate qubits between which the CNOT gate can
be directly applied.  }}
	\label{fig1}
\end{figure}
The calibration parameters of  IBM Q Valencia   on  19 January 2021 are  presented on Table \ref{trr1} \cite{kk}.

\begin{table}[h]
\begin{center}
\caption{The calibration parameters of  IBM Q Valencia on 19 January 2021}
\begin{tabular}{ c c c c c c }
       & $Q_0$ &  $Q_1$ & $Q_2$ & $Q_3$ & $Q_4$ \\
 Readout error ($10^{-2}$) & 4.33 & 2.92  & 6.50 & 2.24 &1.61  \\
Gate error ($10^{-4}$)& 4.35 & 3.14 &10.98 & 6.17 & 9.90 \\

CNOT error  ($10^{-3}$) & CX0$\_$1& CX1$\_$0 &  CX1$\_$2 & CX1$\_$3& CX2$\_$1  \\
                            & 7.70 & 7.70& 13.70 &  12.37 & 13.70 \\
                            &CX3$\_$1 & CX3$\_$4 & CX4$\_$3 &&\\
                            & 12.37 & 23.68 & 23.68 & &\\
\label{trr1}
\end{tabular}
\end{center}
\end{table}

Taking into account the structure of  IBM Q Valencia device, let us examine the graph state defined as
 \begin{eqnarray}
 \mid\psi\rangle=e^{-\frac{itJ}{\hbar}(\sigma_0^x\sigma_1^x+\sigma_1^x\sigma_2^x+\sigma_1^x\sigma_3^x+\sigma_3^x\sigma_4^x)}\mid00000\rangle, \label{q3}
 \end{eqnarray}
which can be associated with graph with structure corresponding to the structure of IBM Q Valencia quantum computer (see Fig. \ref{fig1}).
In the graph  the maximal vertex degree is equal to 3. This is degree of vertex labeled as 1, $\textrm{deg}(V_1)=3$. The the minimal vertex degree is qual to 1 ($\textrm{deg}(V_0)=\textrm{deg}(V_2)=\textrm{deg}(V_4)=1$). For vertex 3 we have $\textrm{deg}(V_3)=2$.
So, according to result (\ref{res}) obtained analytically in the previous section the entanglement of spin labeled as $1$ (the spin corresponds to $q [1]$) with other spins in the graph state (\ref{q3}) reads $E_1(\vert\psi\rangle)=(1-|\cos^{3}\varphi|)/2$. For the entanglement of spin $0$  corresponding  to $q [0]$  with other spins in (\ref{q3}) we have $E_0(\vert\psi\rangle)=(1-|\cos\varphi|)/2$. The same expression we obtain for the entanglement of spin $2$ and spin $4$, namely  $E_2(\vert\psi\rangle)=E_4(\mid\psi\rangle)=E_0(\vert\psi\rangle)$. And the entanglement of spin $3$ (the spin corresponds to $q [3]$) with other spins in the graph state (\ref{q3}) is the following $E_3(\vert\psi\rangle)=(1-\cos^{2}\varphi)/2$.

The quantum protocol for preparing graph state (\ref{q3}) is presented on Fig. \ref{q3}

\begin{figure}[h!]
	\centering
	\includegraphics[width=14cm]{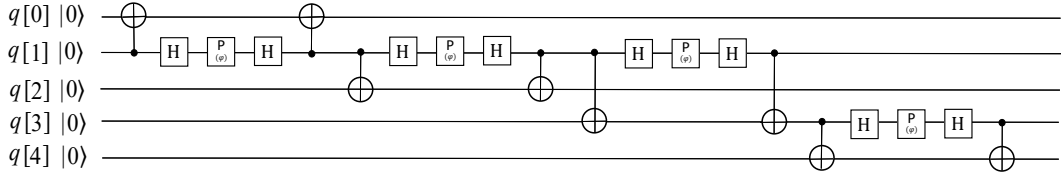}
	\caption{\footnotesize{Quantum protocol for preparing graph state (\ref{q3}).}}
	\label{protok}
\end{figure}
Writing protocol Fig. \ref{q3}, we take into account the calibration parameters of  IBM Q Valencia. Namely, we use representation   $CX_{ij}H_iP_i(2Jt/\hbar)H_iCX_{ij}$ or $CX_{ji}H_jP_j(2Jt/\hbar)H_jCX_{ji}$ for the operator  $\exp({-itJ\sigma_i^x\sigma_j^x}/\hbar)$  dependently on the errors for the quantum bits $q[i]$, $q[j]$. For instance, because the gate error for $q[1]$ is less than gate error for q[0] (see Table \ref{trr1}) to realize operator $\exp({-itJ\sigma_0^x\sigma_1^x}/{\hbar})$ we use the following representation   $CX_{10}H_1P_1(2Jt/\hbar)H_1CX_{10}$ (see first five gates on Fig. \ref{protok}).

\begin{figure}[h!]
	\centering
	\includegraphics[width=3cm]{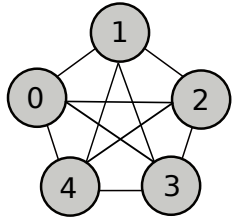}
	\caption{\footnotesize{Graph associated with state (\ref{grf}).}}
	\label{complete_graph}
\end{figure}

\begin{figure}[h!]
	\centering
	\includegraphics[width=14cm]{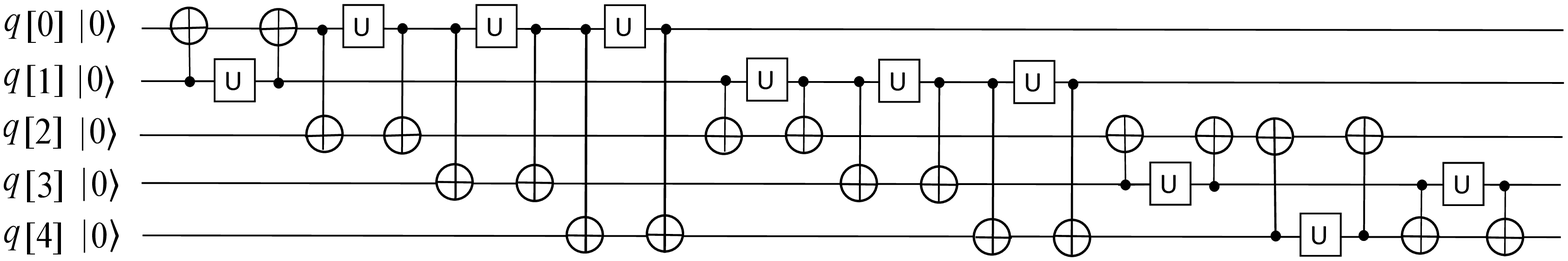}
	\caption{\footnotesize{Quantum protocol for preparing graph state (\ref{grf}). Here $U$ represents gates $HP(\varphi)H$. }}
	\label{comp_prot}
\end{figure}

We have also prepared the graph state
 \begin{eqnarray}
 \mid\psi\rangle=e^{-\frac{itJ}{2\hbar}\sum^{4}_{i=0}\sum^{4}_{j=0}\sigma_i^x\sigma_j^x}\mid00000\rangle,\label{grf}
 \end{eqnarray}
associated with the complete graph (see Fig. \ref{complete_graph}) on IBM Q Valencia quantum computer.
The degree of all vertexes in the graph is 4. Therefore, on the basis of result (\ref{res}) the entanglement of spin l (l=(0..4)) with other spins in the state (\ref{grf}) reads
$E_l(\mid\psi\rangle)=(1-\cos^{4}\varphi)/2$. The quantum protocol for preparing graph state  (\ref{grf}) is presented on Fig. \ref{comp_prot}.

\begin{figure}[!!h]
\subcaptionbox{\label{ff1}}{\includegraphics[scale=0.4, angle=0.0, clip]{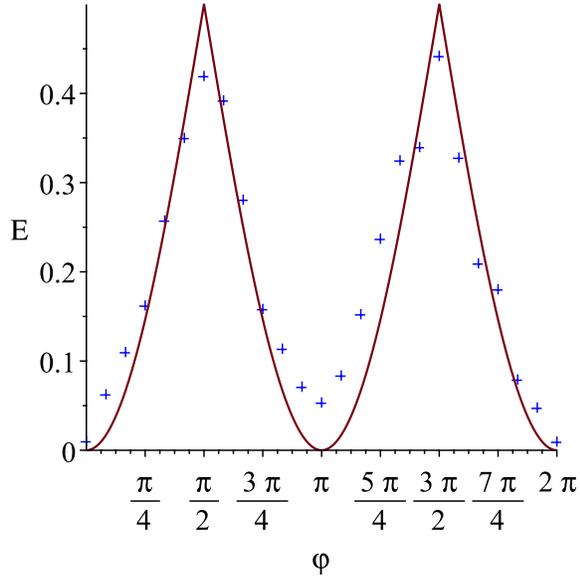}}
\subcaptionbox{\label{ff2}}{\includegraphics[scale=0.4, angle=0.0, clip]{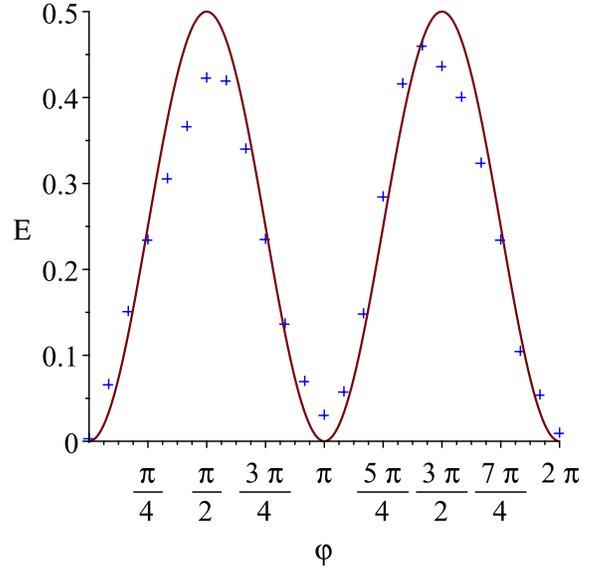}}
\subcaptionbox{\label{ff3}}{\includegraphics[scale=0.4, angle=0.0, clip]{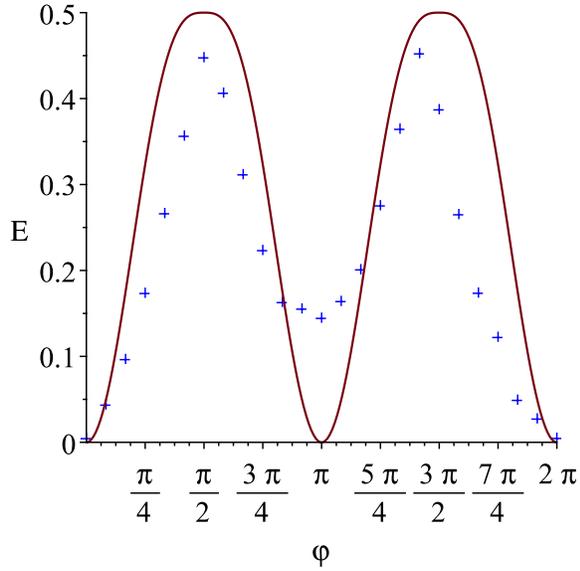}}
\subcaptionbox{\label{ff4}}{\includegraphics[scale=0.4, angle=0.0, clip]{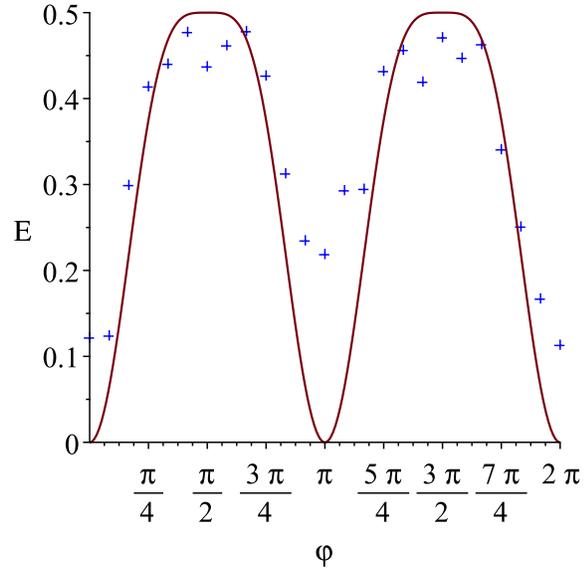}}
\caption{Results of quantifying geometric measure of entanglement on IBM Q Valencia  device  (marked by crosses) and analytical results (line) for spin 4 (a) and spin 3 (b)  spin 1 (c) with other spins in the state (\ref{q3}). Results of quantifying geometric measure of entanglement on IBM Q Valencia device  (marked by crosses) and analytical result (line) for spin 1 with other spins in the state (\ref{grf})  for different values of $\varphi$ (d).}
\label{rr}
\end{figure}

Using  protocols for preparing graph states Fig. \ref{protok}, Fig. \ref{comp_prot} and protocols for  measuring mean values of Pauli operators $\sigma_l^x$, $\sigma_l^y$, $\sigma_l^z$  presented in \cite{Kuzmak} we quantify the geometric measure of entanglement of spins in the states (\ref{q3}) and (\ref{grf})  on the IBM Q Valencia device.
 The entanglement of spins  corresponding to vertexes with degrees  1, 2, 3, 4  was found. Namely, the entanglement of spins labeled by indexes 1, 3, 4 (corresponding qubits are $q [1]$, $q [3]$, $q [4]$) with other spins in the state (\ref{q3}) was calculated. The qubit $q [4]$ was chosen because of small readout error for this qubit in comparison with the errors for $q [0]$ and $q [2]$ (see Table \ref{trr1}). Also the entanglement of spin  1 (corresponding qubit is $q [1]$) with other spins in the graph state (\ref{grf})  was obtained. The qubit $q [1]$ was chosen because of small readout error for this qubit in comparison with other qubits (see Table \ref{trr1}). The results are presented on Fig. \ref{rr}.

Note that for
geometric measure of entanglement of spin 3 and spin 4 with other spins in the graph state (\ref{q3})   we obtained good agreement of experimental results obtained on quantum computer with theoretical ones. Because of redout error and gate error the results for entanglement of spin 1 with other spins in the graph state (\ref{q3})  obtained on quantum computer are not in so good agreement with theoretical ones.   Also, the results of quantifying  geometric measure of entanglement of spin 1 with other spins in the  state (\ref{grf}) corresponding to the complete graph are not in so good agreement with theoretical ones (see Fig. \ref{rr} (d)) as results for entanglement of  spins in the state (\ref{q3}) (see Fig. \ref{rr} (a), (b), (c)) because in quantum protocol  Fig. \ref{comp_prot} more gates are used as in Fig. \ref{protok}, this leads to accumulation of errors.

\section{Conclusions}

The states of many spin systems generated by operator of evolution with Ising Hamiltonian have been considered (\ref{state}). These states  can be associated with graphs with vertexes represented by spins and edges corresponding to interaction between the spins.
The geometric measure of entanglement of a spin with other spins has been found for states associated with graphs with arbitrary adjacency matrixes. We have obtained that the entanglement is the same in the case of antiferromagnetic interaction and ferromagnetic interaction of spins in the system. We have also concluded that the geometric measure of entanglement of a spin with other spins in the graph state is related with graph properties. Namely it depends on the degree of vertex which represents the spin in the graph (\ref{ent}).

Entanglement of a spin with other spins in the graph state (\ref{state}) has been  also quantified on quantum computer.   The state associated with graph which structure corresponds to the structure of  IBM Q Valencia  device (see (\ref{q3}), Fig. \ref{fig1}) and the state corresponding to the complete graph (see (\ref{grf}), Fig. \ref{complete_graph}) have been prepared.
We have quantified the geometric measure of entanglement of spins corresponding to the vertex with degrees 1, 2, 3, in graph  with structure of IBM Q Valencia  quantum computer Fig. \ref{fig1}. Also, the geometric measure of entanglement of a spin  with other spins in the graph state represented by complete graph (\ref{grf}) has been measured. The results obtained on the quantum computer are in good agreement with theoretical one (see Fig. \ref{rr}).

\section*{Acknowledgments}
This work was supported by Project 2020.02/0196  (No. 0120U104801) from National Research Foundation of Ukraine.

\end{document}